\documentstyle[aps,epsf]{revtex}
\tighten
\begin{document}
\draft

\title{Spin-orbit-induced correlations of the local density of states
in two-dimensional electron gas}

\author{V. M. Apalkov$^1$, M. E. Raikh$^1$, and B. Shapiro$^2$ \\
$^1$Department of Physics, University of Utah, Salt Lake City,
UT 84112, USA \\
$^2$Department of Physics, Technion-Israel Institute of
Technology, Haifa 32000, Israel}
\maketitle

\maketitle
\begin{abstract}
We study the local density of  states (LDOS)
of  two-dimensional electrons  in the presence of spin-orbit (SO)
coupling. Although SO coupling has no effect
on the average  density of states,
it manifests itself in  the  correlations of the LDOS. Namely, the
correlation function  acquires
two satellites centered at energy difference equal to
the SO  splitting, $ 2\omega_{\mbox{\tiny SO}}$,
of the electron Fermi surface. For a smooth disorder
the satellites are well separated from the main peak.
  Weak Zeeman splitting
$\omega_{\mbox{\tiny Z}} \ll\omega_{\mbox{\tiny SO}}$
in a parallel magnetic field causes an anomaly in the
shape of the satellites.
We consider the effect of SO-induced satellites in the LDOS
correlations on
the shape of the correlation function of resonant-tunneling conductances
at different source-drain biases, which can be measured experimentally.
This shape is strongly sensitive to the relation between
$\omega_{\mbox{\tiny SO}}$ and $\omega_{\mbox{\tiny Z}}$.
\end{abstract}



\section{Introduction.}
\label{Introduction}

\large

The density of states, $\nu_0$, in 2D electron gas is
energy-independent. It remains energy-independent
in a parallel magnetic field, which causes
the spin-splitting, $2\omega_{\mbox{\tiny Z}}$,
of the electron spectrum. The density of states, corresponding to
each spin branch, is equal to $\nu_0/2$.
Spin-orbit (SO) coupling also results in the splitting of the
electron spectrum into two branches\cite{bychkov84}.
In this case the density of states in each branch, $\mu=\pm 1$,
depends on energy

\begin{equation}
\label{nonperturbed}
\nu_{\mu}(\epsilon)=
\frac{\nu_0}{2}
\Bigl[1+\mu\frac{\omega_{\mbox{\tiny SO}}}
{2\epsilon}\Bigr],
\end{equation}
where $\pm\omega_{\mbox{\tiny SO}}$ is the splitting of
a state which had the energy
$\epsilon \gg \omega_{\mbox{\tiny SO}}$
in the absence of the SO-coupling.
As can be seen from Eq. (\ref{nonperturbed}),
the net density of states is still identically equal to $\nu_0$.
Since the magnitude of the splitting, $2\omega_{\mbox{\tiny SO}}$,
at the Fermi level $\epsilon = E_{\mbox{\tiny F}}$ is inversely
proportional to the length of the spin rotation, this magnitude
represents an  important characteristics
of the 2D structure, containing the electron gas.
The importance of the spin-rotation length was appreciated
since 1990, when the proposal for device application of the
spin-polarized currents was put forward\cite{datta90}. This proposal
has lately attracted a lot of interest.

  If the magnitude of the SO-splitting
is large enough, it can be inferred from the beating pattern
of the Shubnikov-de Haas oscillations, as was first demonstrated by
Dorozhkin and  Olshanetskii\cite{dorozhkin87} for $Si$-based
structures, and subsequently\cite{luo90} by Luo {\em et al.}
for narrow-gap heterostructures. While for large 
$\omega_{\mbox{\tiny SO}}$  (several {\em meV}), 
as in  \cite{luo90}, the beats of the
Shubnikov-de Haas oscillations yield a rather accurate value of
SO-splitting, extracting small splittings
from the beating pattern is complicated in two regards:
(i) very low magnetic fields (with cyclotron energy smaller than
$\omega_{\mbox{\tiny SO}}$)  and, correspondingly, very
low temperatures
are required to observe the beatings; (ii) presence
of even moderate disorder suppresses the
Shubnikov-de Haas oscillations by the Dingle factor, which gets
large at low fields.

In the present paper we demonstrate that the disorder can, actually,
{\em reveal} the SO-splitting even if $\omega_{\mbox{\tiny SO}}$
is smaller than
the single particle scattering rate, $\tau^{-1}$. 
In a disordered sample the local density of states (LDOS) fluctuates 
randomly in space and these fluctuations contain information about 
SO coupling. Our main point is that this information, which is lost
in the average LDOS, is preserved in the correlation function of
LDOS at two different energies, $P(\epsilon_1,\epsilon_2)$.
This function, whose magnitude is inversely proportional to the
disorder, exhibits peaks 
at $(\epsilon_1-\epsilon_2) =\pm 2 \omega_{\mbox{\tiny SO}}$.
Even for $\omega_{\mbox{\tiny SO}}\tau \ll 1$ the peaks
are well pronounced when the disorder is smooth,
so that the transport relaxation time, $\tau_{tr}$,  is much longer
than $\tau$. This is illustrated in Fig.~1. Smooth disorder
implies that the momentum transfer in a single scattering act is
small compared to the Fermi momentum,  $k_{\mbox{\tiny F}}$. In this
way, the states with energies $\epsilon + \omega_{\mbox{\tiny SO}}$
and $\epsilon - \omega_{\mbox{\tiny SO}}$ remain correlated after
many, $\sim \tau_{tr}/\tau$, scattering acts. Therefore,
the SO-induced satellites in $P(\epsilon_1,\epsilon_2)$ are sharp
even in the presence of a strong disorder,
$\tau^{-1} >\omega_{\mbox{\tiny SO}}$, if the condition
$\omega_{\mbox{\tiny SO}}\tau_{tr} \gg 1$ is met. In terms of
underlying physics,  the observation
that a small SO-splitting can be resolved in a smooth potential
 is quite analogous to the observation made in
Ref. \onlinecite{rudin98} concerning the manifestation
of the Landau levels in $P(\epsilon_1,\epsilon_2)$ in the
presence of a strong disorder, when the Landau level structure
in the average density of states is completely smeared 
out\cite{raikh93}.  Let us emphasize, though, that in our case
(unlike Ref.\onlinecite{rudin98}) the density of states in the
absence of disorder is constant, which does not contain 
$\omega_{\mbox{\tiny SO}}$. Thus, it is only in the presence
of disorder that LDOS becomes sensitive to the SO splitting.


Concerning the experimental consequences of SO-induced satellites
in  $P(\epsilon_1,\epsilon_2)$, we note that the
local density of states governs the probability of electron tunneling
from the impurity in the barrier into the
2D gas\cite{prigodin91,lerner92,falko97'}.
This fact was recently utilized in the resonant-tunneling
spectroscopy\cite{falko96,falko97,falko00,falko01,falko01'} of the
LDOS. More specifically, the correlator of the LDOS
determines the behavior of correlation function, $C(\delta V)$, of the
fluctuations of the tunneling conductance
with the change, $\delta V$, of  the source-drain bias.
From the analysis of $C(\delta V)$,
 measured experimentally, the authors\cite{falko01} were able to
extract the quantitative information about the structure of states
and their inelastic lifetime in the disordered emitter.
In the present paper we demonstrate that the SO-induced
 correlations in the LDOS give rise to the new characteristic
features in  $C(\delta V)$. We also demonstrate that these
new features are extremely sensitive to a weak parallel magnetic
field.

The paper is organized as follows.  In Sect. II
we derive the analytic expression for $P(\epsilon_1-\epsilon_2)$
in the case of a smooth disorder. In Sect. III
we give the qulitative explanation for the ratio
between the amplitudes of the main peak in
$P(\epsilon_1-\epsilon_2)$ and of the SO-induced satellites.
In Sect. IV
we use the  shape of  $P(\epsilon_1-\epsilon_2)$  to calculate
the SO-induced features in the correlator, $C(\delta V)$, of the
tunneling conductances. Concluding remarks are presented in Sect. V.

\section{Calculation of $P(\epsilon_1-\epsilon_2)$}
\label{Calculation}
\subsection{Hamiltonian and eigenfunctions}
We choose the conventional form, $\alpha (\hat{\bbox{\sigma }} 
  \times \bbox{k}) \bbox{n}$  ,
 of the SO-term originating from the
confinement potential asymmetry\cite{bychkov84}.
Here $\alpha$ is the coupling constant, $\hat{\bbox{\sigma}}$
is the spin operator, and $\bbox{n}$ is the unit vector
normal to the two-dimensional plane.
In a parallel magnetic field, that induces the Zeeman splitting
$2\omega_{\mbox{\tiny Z}}$, the Hamiltonian of a free electron
is
\begin{equation}
\label{simplest}
\hat{H} = \frac{\hbar ^2 k^2}{2m} + \alpha (\hat{\bbox{\sigma }}
  \times \bbox{k}) \bbox{n} + \omega_{\mbox{\tiny Z}} 
     \hat{\sigma} _x,
\end{equation}
where $m$ is the electron mass.
The eigenfunctions of the
Hamiltonian Eq.~(\ref{simplest}) are classified according to the
chirality, $\mu=\pm 1$, and form two branches of the spectrum,
as illustrated in Fig. 1.  In the vicinity of the Fermi
surface the spectrum, $E_{\mu}(\bbox{k})$, can be simplified
\begin{equation}
\label{vicinity}
E_{\mu } (k) = E_{F} +\epsilon_{\mu }(\bbox{k}),
\end{equation}
 where $\epsilon_{\mu }(\bbox{k})$ is defined as
\begin{equation}
\epsilon _{\mu } (\bbox{k}) = \hbar v_{\mbox{\tiny F}} 
         (k - k_{\mbox{\tiny F}} )  +
\mu \Omega (\bbox{k}) ,
\label{ep_1}
\end{equation}
and
\begin{equation}
\Omega (\bbox{k}) = \sqrt{ \omega _{\mbox{\tiny SO}} ^2 
        +\omega_{\mbox{\tiny Z}}^2 +
 2  \omega _{\mbox{\tiny SO}}  \omega_{\mbox{\tiny Z}} 
         \sin \phi _{\bbox{k}}} .
\label{ep_2}
\end{equation}
Here $v_{\mbox{\tiny F}} = \hbar k_{\mbox{\tiny F}}/m$ is the 
Fermi velocity,
$ \omega _{\mbox{\tiny SO}}=\alpha k_{\mbox{\tiny F}}$, and
 $\phi _{\bbox{k}}$ is the azimuthal angle of $\bbox{k}$.
It is convenient to introduce the projection operators
$\hat{\Lambda }_{\mu}(\bbox{k})$, which are defined through
the eigenfunctions
\begin{equation}
\label{eigen}
\chi_{\mu}(\bbox{k})=  \frac{1}{2^{1/2}} \left( \begin{array}{c}
   1 \\  -i\mu \exp (i \varphi _{\bbox{k}} )
 \end{array} \right),
\end{equation}
of the Hamiltonian Eq. (\ref{simplest}) as
$\hat{\Lambda }_{\mu}(\bbox{k})=
\chi_{\mu}(\bbox{k})\chi_{\mu}^{\dagger}(\bbox{k})$,
so that\cite{chen99'}
\begin{equation}
\hat{\Lambda } _{\mu } ( \bbox{k} ) =
 \frac{1}{2} \left( \begin{array}{cc}
   1 &   i\mu \exp (-i \varphi _{\bbox{k}} )  \\
  -i\mu \exp (i \varphi _{\bbox{k}} ) & 1
 \end{array} \right),
 \label{lambda1}
\end{equation}
where the angle $\varphi _{\bbox{k}}$ is related to the
azimuthal angle $\phi _{\bbox{k}}$ as
\begin{equation}
\tan \varphi _{\bbox{k}} =\tan \phi _{\bbox{k}} +
\frac{\omega_{\mbox{\tiny Z}}}{ \omega
  _{\mbox{\tiny SO}}\cos \phi _{\bbox{k}}} .
\label{angle}
\end{equation}
In terms of projection operators the Hamiltonian
 can be presented in a simple form
$\label{hamil}
\hat{H}=\sum _{\mu }
E_{\mu } (\bbox{k}) \hat{\Lambda } _{\mu } (\bbox{k})$.

\subsection{Correlator of the LDOS in the chirality representation}
In the presence of disorder,
the LDOS at energy $\epsilon$ at point $\bbox{r}$  is
defined as
\begin{equation}
\nu (\bbox{r}, \epsilon ) = \frac{1}{2\pi i} \mbox{Tr} \left[
      \hat{G}^{A}(\bbox{r},\bbox{r},\epsilon ) -
      \hat{G}^{R}(\bbox{r},\bbox{r},\epsilon )       \right],
\end{equation}
where the advanced and retarded Green's functions are $2\times 2$
matrices in the chirality space, and the trace is taken over
different chiralities.  The main
contribution to the LDOS correlator comes from the cross terms
$G^{(R)}G^{(A)}$
\begin{equation}
\label{P}
P (\epsilon -\epsilon^{\prime}) =
  \frac{1}{\nu_0^2} \langle \delta \nu (\bbox{r},\epsilon )
   \delta \nu (\bbox{r},\epsilon^{\prime })\rangle
 =
 \frac{1}{2\pi^2 \nu_0^2 } \mbox{Re} \langle
 \mbox{Tr}  \hat{G}^{R}(\bbox{r},\bbox{r},\epsilon )
 \mbox{Tr}  \hat{G}^{A}(\bbox{r},\bbox{r},
 \epsilon ^{\prime })
   \rangle,
\end{equation}
where $\delta \nu (\bbox{r},\epsilon )$ is the deviation
of the LDOS from the disorder-free value, $\nu_0$. It is
easy to see from Eq. (\ref{ep_1}) that $\nu_0 = m/\pi\hbar^2$,
and does not depend on energy.
To evaluate the correlator Eq. (\ref{P}) it is convenient to
use the chirality representation of the Green functions
\begin{equation}
\hat{G}^{R,A}
(\epsilon , \bbox{p})
 =  \! \! \sum _{\mu }
   \frac{ \hat{\Lambda } _{\mu } (\bbox{p}) }
        { \epsilon - \epsilon _{\mu } (\bbox{p}) \pm  \frac{i}{2\tau } }
        = \! \! \sum _{\mu }
    \hat{\Lambda } _{\mu } (\bbox{p})
     G^{R,A} _{\mu } (\epsilon , \bbox{p}) ,
\label{green}
\end{equation}
where $\tau$ is the scattering time.
As in the absence of the SO-coupling, the LDOS correlator represents the
sum of the diffuson and cooperon contributions\cite{prigodin93}.
Generalized expressions for the corresponding contributions in the
chirality representation read
\begin{eqnarray}
P_D (\epsilon -\epsilon ^{\prime } )
  & = & \frac{1}{2\pi ^2 \nu_0^2 }
\mbox{Re}
 \int \frac{d \bbox{q}}{(2\pi )^2}
\int  \frac{d \bbox{p}}{(2\pi )^2}
\int  \frac{d \bbox{p}^{\prime }}{(2\pi )^2}
\sum _{\{\mu , \eta \}}
\mbox{\Large $\Gamma$} ^{\mu_1 \eta_1 }_{\mu_2 \eta_2 }
 (\bbox{p}, \bbox{p}^{\prime }, \bbox{q}, \epsilon - \epsilon ^{\prime })
\Phi_{\mu_1, \eta_1}^{\ast }(\bbox{p},\bbox{p}^{\prime})
  \Phi_{\mu_2, \eta_2}(\bbox{p},\bbox{p}^{\prime})
  \nonumber \\
& & \times  G_{\mu_1}^{R} (\epsilon , \bbox{p}+\bbox{q})
 G_{\eta_1}^{R} (\epsilon , \bbox{p}^{\prime }+\bbox{q})
 G_{\mu _2}^{A} (\epsilon^{\prime } , \bbox{p})
 G_{\eta_2}^{A} (\epsilon^{\prime } , \bbox{p}^{\prime }),
 \label{D_1}
\end{eqnarray}

\begin{eqnarray}
P_C (\epsilon -\epsilon ^{\prime } )
  & = & \frac{1}{2\pi ^2 \nu_0^2 }
\mbox{Re}
 \int \frac{d \bbox{q}}{(2\pi )^2}
\int  \frac{d \bbox{p}}{(2\pi )^2}
\int  \frac{d \bbox{p}^{\prime }}{(2\pi )^2}
\sum _{\{\mu , \eta \}}
\mbox{\Large $\Gamma$} ^{\mu_1 \eta_1 }_{\mu_2 \eta_2 }
 (\bbox{p}, \bbox{p}^{\prime }, \bbox{q}, \epsilon - \epsilon ^{\prime })
\Phi_{\mu_1, \eta_1}(\bbox{p},\bbox{p}^{\prime})
  \Phi_{\mu_2, \eta_2}^{\ast }(\bbox{p},\bbox{p}^{\prime})
  \nonumber \\
& & \times  G_{\mu_1}^{R} (\epsilon , \bbox{q} - \bbox{p})
 G_{\eta_1}^{R} (\epsilon , \bbox{q} - \bbox{p}^{\prime })
 G_{\mu _2}^{A} (\epsilon^{\prime } , \bbox{p})
 G_{\eta_2}^{A} (\epsilon^{\prime } , \bbox{p}^{\prime }) =
P_D (\epsilon ^{\prime }-\epsilon  ),
 \label{C_1}
\end{eqnarray}
where
\begin{equation}
\Phi_{\mu, \eta}(\bbox{p},\bbox{p}^{\prime}) =
\chi_{\mu}^{\dagger } (\bbox{p}) \chi_{\eta}
(\bbox{p}^{\prime}) = \frac{1}{2}\Bigl\{1+\mu \eta \exp\left(
i[\varphi _{\bbox{p}}- \varphi _{\bbox{p}^{\prime}}]
\right) \Bigr\}.
\end{equation}
The two-particle vertex functions 
$\mbox{\Large $\Gamma$} ^{\mu_1 \eta_1 }_{\mu_2 \eta_2 }
 (\bbox{p}, \bbox{p}^{\prime }, \bbox{q}, \epsilon)$ in
Eqs. (\ref{D_1}),(\ref{C_1}) satisfy the matrix Dyson-type equation
\begin{equation}
\Gamma ^{\mu_1 \eta_1}_{\mu_2 \eta_2 }
 (\bbox{p}, \bbox{p}^{\prime }, \bbox{q}, \epsilon ) =
S\mbox{\Large $($} \bbox{p}-\bbox{p} ^{\prime }
  \mbox{\Large $)$}
  \Phi_{\mu_1, \eta_1}(\bbox{p},\bbox{p}^{\prime})
  \Phi_{\mu_2, \eta_2}(\bbox{p},\bbox{p}^{\prime})
 +  \! \! \!  \int \! \! \frac{d \bbox{p}_1}{(2\pi )^2}
 \sum_{\xi_1, \xi_2 }
K^{\mu_1 \xi_1}_{\mu_2 \xi_2} (\bbox{p}, \bbox{p}_1,\bbox{q},\epsilon )
  \Gamma ^{\xi_1 \eta_1}_{\xi_2 \eta_2 }
  (\bbox{p}_1, \bbox{p}^{\prime }, \bbox{q}, \epsilon )
\label{dyson0}
\end{equation}
 with a kernel
\begin{equation}
 K^{\mu_1 \xi_1}_{\mu_2 \xi_2}(\bbox{p}, \bbox{p}_1,\bbox{q},\epsilon )  =
 \Phi_{\mu_1, \xi_1}(\bbox{p},\bbox{p}^{\prime})
  \Phi_{\mu_2, \xi_2}(\bbox{p},\bbox{p}^{\prime})
  G_{\xi_1}^{R} (\omega +\epsilon , \bbox{p}_1+\bbox{q})
 G_{\xi_2}^{A} (\omega , \bbox{p}_1)
S\mbox{\Large $($} \bbox{p}-\bbox{p} ^{\prime }
  \mbox{\Large $)$},
\label{kernel0}
\end{equation}
where the function $S\mbox{\Large $($} \bbox{p}-\bbox{p} ^{\prime }
  \mbox{\Large $)$}$ is the Fourier transform of the correlator
of the  random potential. Upon integration over $\vert\bbox{p}_1\vert$ 
Eq. (\ref{dyson0}) takes the form
\begin{equation}
\Gamma ^{\mu_1 \eta_1}_{\mu_2 \eta_2 }
 (\bbox{p}, \bbox{p}^{\prime }, \bbox{q}, \epsilon ) =
S\mbox{\Large $($} \bbox{p}-\bbox{p} ^{\prime }
  \mbox{\Large $)$}
  \Phi_{\mu_1, \eta_1}(\bbox{p},\bbox{p}^{\prime})
  \Phi_{\mu_2, \eta_2}(\bbox{p},\bbox{p}^{\prime})
 +  \! \! \!  \int \! \! \frac{d \phi_{\bbox{p}_1}}{(2\pi )^2}
 \sum_{\xi_1, \xi_2 }
\tilde{K}^{\mu_1 \xi_1}_{\mu_2 \xi_2} 
  (\bbox{p}, \bbox{p}_1,\bbox{q},\epsilon )
  \Gamma ^{\xi_1 \eta_1}_{\xi_2 \eta_2 }
  (\bbox{p}_1, \bbox{p}^{\prime }, \bbox{q}, \epsilon )
\label{dyson}
\end{equation}
where we have introduced the modified kernel
 $\tilde{K}^{\mu_1 \xi_1}_{\mu_2 \xi_2}$, defined as
\begin{equation}
\tilde{K}^{\mu_1 \xi_1}_{\mu_2 \xi_2}(\bbox{p}, \bbox{p}_1,\bbox{q},\epsilon )  =  \left(\frac{m \tau }{2 \pi }\right)
\frac{
 \Phi_{\mu_1, \xi_1}(\bbox{p},\bbox{p}^{\prime})
  \Phi_{\mu_2, \xi_2}(\bbox{p},\bbox{p}^{\prime})
S\mbox{\Large $($} \bbox{p}-\bbox{p} ^{\prime }
  \mbox{\Large $)$}}{
1 - i \mbox{\Large $($} \omega -\xi_1\Omega (\bbox{p}_1)
       +\xi_2 \Omega (\bbox{p}_1+\bbox{q}) \mbox{\Large $)$} \tau
 +     i \hbar q v_{\mbox{\tiny F}}   \tau \cos(\phi _{\bbox{p}} -
      \phi _{\bbox{q}})
},
\label{kernel}
\end{equation}

When the random potential is smooth, the function
$S$ restricts the difference $\bbox{p}-\bbox{p}^{\prime}$ within a
narrow domain $\vert\bbox{p}-\bbox{p}^{\prime}\vert
\ll k_{\mbox{\tiny F}}$. As we will see below this leads to a
drastic simplification of the system Eqs. (\ref{dyson}), (\ref{kernel}).

\subsection{Smooth potential; $|\epsilon_1-\epsilon_2|
\approx 2 \omega_{SO}$}

In the case of a smooth potential the factors
$\Phi_{\mu, \eta}(\bbox{p},\bbox{p}^{\prime})$ 
for coinsiding chiralities, $\mu = \eta$, differ strongly from 
those with opposite chiralities, $\mu=-\eta$.  
Indeed, when $\vert \bbox{p}-\bbox{p}^{\prime}\vert$ is
small compared to $k_{\mbox{\tiny F}}$, we have
$\phi_{\bbox{p}} \approx \phi_{\bbox{p}^{\prime}}$ and thus
 $\varphi_{\bbox{p}} \approx \varphi_{\bbox{p}^{\prime}}$, so that
$\Phi_{\mu,-\mu} \ll 1$, whereas $\Phi_{\mu,\mu}$ is close to unity
and can be presented as
\begin{equation}
\label{factor}
\Phi_{\mu,\mu}(\bbox{p},\bbox{p}^{\prime})= 1-
  \frac{\left(\varphi_{\bbox{p}}- \varphi_{\bbox{p}^{\prime}}\right)^2
 }{8} .
\end{equation}
Using Eq. (\ref{angle}), the
factor $\Phi_{\mu,\mu}$ can be expressed through
the angle, $\phi_{\bbox{p}}-\phi_{\bbox{p}^{\prime}}$,  between the vectors $\bbox{p}$
and $\bbox{p}^{\prime}$ as follows
\begin{equation}
\Phi_{\mu,\mu}(\bbox{p},\bbox{p}^{\prime})= 1-\frac{\omega _{\mbox{\tiny SO}}^2
  ( \omega _{\mbox{\tiny SO}} +\omega _{\mbox{\tiny Z}} \sin \phi _{\bbox{p}} )^2 }
  {4 \Omega ^4(\bbox{p}) }
    \mbox{\Large $($} \phi _{\bbox{p}^{\prime }} - \phi _{\bbox{p}}  \mbox{\Large $)$}^2 .
\label{lambda_1}
\end{equation}
The latter simplification allows to set $\Phi_{\mu,-\mu}=0$ and
$\Phi_{\mu,\mu}=1$ everywhere except for the kernel of the
Dyson equation. As a result, the system Eqs. (\ref{dyson}) gets
decoupled into {\em closed} equations for the elements,
$\Gamma ^{\mu,\mu}_{\eta,\eta}$ with $\mu \neq \eta$,  of
the matrix $\Gamma ^{\mu_1 \eta_1}_{\mu_2 \eta_2 }$. The underlying 
reason for this decoupling is that these elements describe the 
two-particle motion, in course of which one particle moves within 
the branch $\mu$ whereas another particle moves within the branch 
$\eta$. For a smooth potential coupling of these elements to the 
other branch-nonconserving elements is small in parameter 
$\left(1-\Phi_{\mu,\mu}\right) \ll 1$. With the above simplification
Eq. (\ref{dyson}) takes the form\cite{apalkov02}
\begin{equation}
\label{vertex}
\Gamma ^{\mu,\mu}_{\eta,\eta}(\bbox{p}, \bbox{p}^{\prime },
\bbox{q}, \omega ) =
S\mbox{\Large $($}
         k_{\mbox{\tiny F}}|\phi _{\bbox{p}} - \phi _{\bbox{p} ^{\prime }}|
  \mbox{\Large $)$}
  +  \! \! \!  \int \! \! \frac{d \phi_{\bbox{p}_1}}{(2\pi )^2}
   \tilde{K}^{\mu,\mu}_{\eta,\eta} 
    (\bbox{p}, \bbox{p}_1,\bbox{q},\omega )
  \Gamma ^{\mu,\mu}_{\eta,\eta}(\bbox{p}_1, \bbox{p}^{\prime }, \bbox{q}, 
   \omega),
\end{equation}
while the kernel Eq. (\ref{kernel}) simplifies to
\begin{equation}
\tilde{K}^{\mu,\mu}_{\eta,\eta}
(\bbox{p}, \bbox{p}_1,\bbox{q},\omega )
\! \! = \! \!            \left(  \frac{m \tau }{2 \pi } \right) \! \!
  \frac{\left| \Phi_{\mu,\mu}(\bbox{p}, \bbox{p}_1)\right|^2 S
 \mbox{\Large $($} k_{\mbox{\tiny F}}|\phi _{\bbox{p}} -
      \phi _{\bbox{p}_1}|\mbox{\Large $)$}  }
   {1 \! -\!  i \tau
  \left[\omega \!- \! \left(\mu-\eta\right)\Omega(\bbox{p})  -
     \hbar q v_{\mbox{\tiny F}} \cos(\phi _{\bbox{p}} - \phi _{\bbox{q}}) \right]}.
 \label{kernel_1}
\end{equation}
Substituting Eq. (\ref{kernel_1}) into Eq. (\ref{vertex}) yields 
an integral equation for the vertex $\Gamma^{\mu,\mu}_{\eta,\eta}$.
Note, that this equation differs from the standard equation for 
the vertex in the absence of the SO-coupling  only by the factor
$\left| \Phi_{\mu,\mu}(\bbox{p}, \bbox{p}_1)\right|^2$
 in the integrand, where 
$\Phi_{\mu,\mu}(\bbox{p}, \bbox{p}_1)$ is defined by 
Eq. (\ref{factor}). Thus, if the small difference 
of this factor from unity is neglected, the solution for  
$\Gamma^{\mu,\mu}_{\eta,\eta}$ would contain a conventional 
diffusive pole at $\omega = (\mu - \eta)\Omega(\bbox{p}) - iDq^2$, 
where $D=v_{\mbox{\tiny F}}^2\tau_{tr}/2$ is the diffusion 
coefficient, and
\begin{equation}
\label{transport}
\tau_{tr}^{-1}=\frac{1}{2}\int \frac{d \bbox{p}}{(2 \pi )^2}
 \left(\phi _{\bbox{k}} - \phi _{\bbox{p}}\right)^2
S \mbox{\Large $($} |\bbox{k} - \bbox{p}| \mbox{\Large $)$}
  \delta \!\left( E_{\mu } (\bbox{k}) -
                  E_{\mu }(\bbox{p}) \right).
\end{equation}
is the transport relaxation time. Taking into account the small 
correction originating from the difference 
$1-\left| \Phi_{\mu,\mu}(\bbox{p}, \bbox{p}_1)\right|^2$  amounts
to the imaginary shift of the pole position by $i\tau_{int}^{-1}$, 
where $\tau_{int}$ is defined as
\begin{equation}
\label{int}
\tau_{int}^{-1}(\bbox{k})=\frac{1}{4}\int \frac{d \bbox{p}}{(2 \pi )^2}
 \left(\varphi _{\bbox{k}} - \varphi _{\bbox{p}}\right)^2
S \mbox{\Large $($} |\bbox{k} - \bbox{p}| \mbox{\Large $)$}
  \delta \!\left( E_{\mu } (\bbox{k}) -
                  E_{\mu }(\bbox{p}) \right).
\end{equation}
The meaning of $\tau_{int}$ is the interbranch  scattering time. 
Indeed, a general expression for the scattering time between 
the branches with different chiralities can be written as
\begin{eqnarray}
\tau _{\mu, -\mu }^{-1}(\bbox{k})  & =  &  \int \frac{d \bbox{p}}{(2 \pi )^2}
 \mbox{Tr} \left( \hat{\Lambda } _{\mu } ( \bbox{k} )
                   \hat{\Lambda } _{-\mu } ( \bbox{p} )  \right)
 S\mbox{\Large $($}
|\bbox{k} - \bbox{p}|  \mbox{\Large $)$}
  \delta \! \left( E_{\mu } (\bbox{k}) -
                  E_{-\mu }(\bbox{p}) \right)  \nonumber \\
& = & \int \frac{d \bbox{p}}{(2 \pi )^2}
 \left( 1- \left| \Phi_{\mu,\mu}(\bbox{k}, \bbox{p})\right|^2  \right)
 S\mbox{\Large $($}
|\bbox{k} - \bbox{p}|  \mbox{\Large $)$}
  \delta \! \left( E_{\mu } (\bbox{k}) -
                  E_{-\mu }(\bbox{p}) \right).
\label{tau_2}
\end{eqnarray}
Comparing Eqs. (\ref{int}) and  (\ref{tau_2}), we see that they 
differ only by the arguments of the $\delta$-functions. This 
difference is negligible when 
$\omega_{\mbox{\tiny SO}}\ll  E_{\mbox{\tiny F}}$, and thus we have
$\tau_{\mu,-\mu}(\bbox{k}) \approx \tau_{int}(\bbox{k})$. 
We also see that the integral Eq. (\ref{int}) in the expression 
for $\tau_{int}^{-1}$ differs from the integral in  
Eq. (\ref{transport}) for $\tau_{tr}^{-1}$ only by the replacement 
$\varphi_{\bbox{k}} \rightarrow \phi_{\bbox{k}}$. Using the 
relation Eq. (\ref{lambda_1}) between the two angles, we can 
express $\tau_{int}$ through the transport relaxation time
\begin{equation}
\tau _{int} (\phi ) =
  2 \tau_{tr} \frac{\left(\omega_{\mbox{\tiny Z}}^2 +
\omega_{\mbox{\tiny SO}}^2 + 2 \omega_{\mbox{\tiny Z}}\omega_{\mbox{\tiny SO}}
 \sin \phi \right)^2}{\omega_{\mbox{\tiny SO}}^2 \left(
 \omega_{\mbox{\tiny SO}}+\omega_{\mbox{\tiny Z}}\sin \phi \right)^2}.
\label{tau5}
\end{equation}
With shifted diffusion pole, the final expression for the 
vertex $\Gamma ^{\mu \mu }_{\eta \eta }$ takes the form
\begin{equation}
\Gamma ^{\mu \mu }_{\eta \eta }
(\bbox{p}, \bbox{p}^{\prime }, \bbox{q}, \omega)
\!\! =
  \!\!  \frac{ i S\mbox{\Large $($}
                 k_{\mbox{\tiny F}}|\phi _{\bbox{p}} - \phi _{\bbox{p} ^{\prime }}|
  \mbox{\Large $)$}}
 {    \tau  \left[   \omega - (\mu -\eta) \Omega (\bbox{p}) +  i D q^2
     + i   \tau ^{-1}_{int}(\bbox{p}) \right] }.
\label{Gamma_0}
\end{equation}
Substituting this form into Eq. (\ref{D_1}) yields the following expression
for the diffuson contribution to the correlator of LDOS
\begin{equation}
P_D(\epsilon ) =  \frac{\tau }{16\pi^3 }~\sum_{\mu\neq \eta}
\mbox{Re} \int  _0 ^{1/v_{\mbox{\tiny F}}\tau_{tr}} dq~q
\int _0^{2 \pi} \frac{d \phi_{\bbox{p}}}{2\pi }
\int _0^{2 \pi} \frac{d \phi_{\bbox{p}^{\prime }}}{2\pi }
\frac{ S\mbox{\Large $($} k_{\mbox{\tiny F}}|\phi _{\bbox{p}} -
                                    \phi _{\bbox{p} ^{\prime }}|\mbox{\Large $)$}}
 {-i\left[\epsilon - (\mu-\eta)\Omega (\bbox{p}) \right] + D q^2
     + \tau ^{-1}_{int}(\bbox{p}) }.
 \label{H_2}
\end{equation}
Two integrations (over $q$ and over $\phi_{\bbox{p}^{\prime }}$) in Eq. (\ref{H_2})
can be readily performed resulting in the following  expression for
SO-induced satellites $P_D(\epsilon )$
\begin{equation}
\label{diff}
P_D(\epsilon ) =  \frac{1}{4\pi E_{\mbox{\tiny F}}\tau_{tr}}~
\sum_{\mu\neq \eta}
{\cal L}_{\mu,\eta}(\epsilon),
\end{equation}
where the functions ${\cal L}_{\mu,\eta}(\epsilon)$ are the 
following azimuthal averages
\begin{equation}
{\cal L}_{\mu,\eta}(\epsilon )= \int _0^{2 \pi} \frac{d \phi }{2\pi }
\ln \left[ \mbox{\Large $($} \epsilon - 
   (\mu-\eta)\Omega (\phi )\mbox{\Large $)$}^2
 \tau_{tr}^2 +
 \frac{\tau_{tr} ^2}{\tau _{int}^2 (\phi) }
  \right].
\label{L}
\end{equation}
Since ${\cal L}_{\mu,\eta}(\epsilon) = {\cal L}_{\eta,\mu}(-\epsilon)$
the cooperon contribution,  $P_C(\epsilon )$, to the LDOS 
correlator coincides with $P_D(\epsilon )$, as it should be 
expected on general grounds. Thus, Eqs. (\ref{diff}) and
(\ref{L}) constitute our final result. The functions
${\cal L}_{\mu,\eta}(\epsilon)$, which
determine the energy dependence of the correlator, exhibit a singular
dependence on a weak
magnetic field, $\omega_{\mbox{\tiny Z}} \ll \omega_{\mbox{\tiny SO}}$
for $\mu = -\eta$, as demonstrated in the next subsection.

\subsection{Shape of the satellites}
As it follows from Eq. (\ref{tau5}),
in a weak magnetic field we have
$\tau_{int}(\phi)\approx 2\tau_{tr}=const(\phi)$.
Then for the satellites, centered at
$\epsilon = \pm 2\omega _{\mbox{\tiny SO}}$ we can identify
the $\phi$-dependence of the integrand in
Eq. (\ref{L}). This dependence
comes from
$\Omega(\phi)$, defined by Eq. (\ref{ep_2}). For
$\omega_{\mbox{\tiny Z}} \ll \omega_{\mbox{\tiny SO}}$, we have
$\Omega(\phi)\approx \omega _{\mbox{\tiny SO}} + \omega _{\mbox{\tiny Z}}\sin\phi$.
Upon substituting this form into Eq. (\ref{L}), the integration can be performed
analytically with the use of the following identity
\begin{equation}
\frac{1}{2\pi }\int_0^{2\pi } d\phi \ln \left(  a+b\sin \phi \right)^2=
\left\{ \begin{array}{ll}  \ln \left[ a+ \sqrt{ a^2 - b^2}\right]^2
   & \mbox{~~for ~}  |a|>|b|  \\
2 \ln |b|  & \mbox{~~for ~}  |a|<|b|
\end{array} \right. ,
\end{equation}
 yielding
\begin{equation}
P (\epsilon ) = -\left( \frac{1}{ \pi E_F \tau _{tr}}  \right)
\left\{ \begin{array}{ll}  \ln
 \left| \epsilon  -  2\omega _{\mbox{\tiny SO}}\right| \tau_{tr}  
     & \mbox{~~for ~}
|\epsilon - 2 \omega _{\mbox{\tiny SO}} | >  \omega_{\mbox{\tiny Z}} \\
 \ln \omega_{\mbox{\tiny Z}} \tau_{tr}  & \mbox{~~for ~}
|\epsilon - 2 \omega _{\mbox{\tiny SO}} | \leq  \omega_{\mbox{\tiny Z}}
\end{array} \right. .
\end{equation}
The  satellites of the correlator $P(\epsilon)$,
calculated numerically without the assumption
$\omega_{\mbox{\tiny Z}} \ll \omega_{\mbox{\tiny SO}}$, are
shown in Fig. 2. 
We see that
the plateau in the shape of the satellites is very well pronounced,
but slightly tilted.
The reason for this tilt is the effect of the ``background'' originating
from the main peak.


\subsection{Smooth potential;
$\vert \epsilon_1-\epsilon_2\vert \ll \omega_{SO}$}

In this subsection we solve the system of Eqs. (\ref{dyson}) 
with kernels defined by Eq.(\ref{kernel})) for small 
energy difference
$\vert \epsilon_1-\epsilon_2\vert \ll \omega_{\mbox{\tiny SO}}$.
Firstly, we point out that the vertex functions
$\Gamma ^{\mu,\mu}_{\mu,\mu}$ and $\Gamma ^{-\mu,\mu}_{-\mu,\mu}$
are relevant, since in Eq. (\ref{kernel}) 
$\xi_1$ must be equal to $\xi_2$.
 At the first glance, in equation (\ref{dyson}) for
  $\Gamma ^{\mu,\mu}_{\mu,\mu}$,  only the term in the r.h.s.
containing the same $\Gamma ^{\mu,\mu}_{\mu,\mu}$ should be kept.
Indeed, the coupling of this term to $\Gamma ^{-\mu,\mu}_{-\mu,\mu}$
 is determined by the kernel $\tilde{K}^{-\mu,\mu}_{-\mu,\mu}$,
which, as it is seen from Eq. (\ref{kernel}) contains a square
of the small parameter $\Phi_{-\mu,\mu}$. However, unlike the case
of the satellites, the ``feedback'' becomes important for
small $\vert \epsilon_1-\epsilon_2\vert$. Namely,  the equation
(\ref{dyson}) for $\Gamma ^{-\mu,\mu}_{-\mu,\mu}$ contains in the r.h.s.
the coupling to  $\Gamma ^{\mu,\mu}_{\mu,\mu}$ with the same
small coefficient $\propto  \Phi_{-\mu,\mu}^2$. Specifics of the
small energy difference, as compared to satellites, is that {\em both}
$\Gamma ^{\mu,\mu}_{\mu,\mu}$ and $\Gamma ^{-\mu,\mu}_{-\mu,\mu}$ are
resonant in this case. Thus Eq. (\ref{dyson}) reduces to
the following system of coupled equations
\begin{eqnarray}
\label{vertex_1}
\Gamma ^{\mu,\mu}_{\mu,\mu}(\bbox{p}, \bbox{p}^{\prime },
\bbox{q}, \omega ) &  = &
S\mbox{\Large $($}
     k_{\mbox{\tiny F}}|\phi _{\bbox{p}} -
  \phi _{\bbox{p} ^{\prime }}| \mbox{\Large $)$}
  +   \int \! \! \frac{d \phi_{\bbox{p}_1}}{(2\pi )^2}
   \tilde{K}^{\mu,\mu}_{\mu,\mu} (\bbox{p}, \bbox{p}_1,\bbox{q},\omega )
  \Gamma ^{\mu,\mu}_{\mu,\mu}(\bbox{p}_1,
  \bbox{p}^{\prime }, \bbox{q}, \omega)   \nonumber \\
 & & +   \int \! \! \frac{d \phi_{\bbox{p}_1}}{(2\pi )^2}
   \tilde{K}^{\mu,-\mu}_{\mu,-\mu} 
(\bbox{p}, \bbox{p}_1,\bbox{q},\omega )
  \Gamma ^{-\mu,\mu}_{-\mu,\mu}(\bbox{p}_1,
  \bbox{p}^{\prime }, \bbox{q}, \omega) ,
\end{eqnarray}
\begin{eqnarray}
\label{vertex_2}
\Gamma ^{-\mu,\mu}_{-\mu,\mu}(\bbox{p}, \bbox{p}^{\prime }, \bbox{q}, \omega )
 & = &
S\mbox{\Large $($}
         k_{\mbox{\tiny F}}|\phi _{\bbox{p}} - \phi _{\bbox{p} ^{\prime }}|
\mbox{\Large $)$}
       \left| \Phi_{\mu,-\mu}(\bbox{p}, \bbox{p}_1)\right|^2
  +  \int \! \! \frac{d \phi_{\bbox{p}_1}}{(2\pi )^2}
   \tilde{K}^{-\mu,-\mu }_{-\mu,-\mu} (\bbox{p}, \bbox{p}_1,\bbox{q},\omega )
  \Gamma ^{-\mu ,\mu}_{-\mu ,\mu}(\bbox{p}_1, \bbox{p}^{\prime }, \bbox{q}, \omega)
 \nonumber \\
& &  +  \int \! \! \frac{d \phi_{\bbox{p}_1}}{(2\pi )^2}
   \tilde{K}^{-\mu,\mu }_{-\mu,\mu} (\bbox{p}, \bbox{p}_1,\bbox{q},\omega )
  \Gamma ^{\mu ,\mu}_{\mu ,\mu}(\bbox{p}_1, \bbox{p}^{\prime }, \bbox{q}, \omega).
\end{eqnarray}
Solution of the  system (\ref{vertex_1}), (\ref{vertex_2})
yields\cite{dmitriev01}
\begin{equation}
\Gamma ^{\mu \mu }_{\mu \mu }
(\bbox{p}, \bbox{p}^{\prime }, \bbox{q}, \omega)
\!\! =           \frac{i}{2\tau }   S\mbox{\Large $($}
 k_{\mbox{\tiny F}}|\phi _{\bbox{p}} - \phi _{\bbox{p} ^{\prime }}|\mbox{\Large $)$}
\left\{
 \frac{ 1 }
 {   \omega  +  i D q^2
     +  i   \tau ^{-1}_{f}(\bbox{p}) } +
  \!\!  \frac{ 1}
 {    \omega  +  i D q^2  + 2 i   \tau ^{-1}_{int}(\bbox{p})  }
   \right\} ,
\label{Gamma_1}
\end{equation}
\begin{equation}
\Gamma ^{-\mu \mu }_{-\mu \mu }
(\bbox{p}, \bbox{p}^{\prime }, \bbox{q}, \omega)
\!\! =           \frac{i}{2\tau }   S\mbox{\Large $($}
 k_{\mbox{\tiny F}}|\phi _{\bbox{p}} - \phi _{\bbox{p} ^{\prime }}|\mbox{\Large $)$}
\left\{
 \frac{ 1 }
 {   \omega  +  i D q^2
     +  i   \tau ^{-1}_{f}(\bbox{p}) } -
  \!\!  \frac{ 1}
 {    \omega  +  i D q^2  + 2 i   \tau ^{-1}_{int}(\bbox{p})  }
   \right\} ,
\label{Gamma_2}
\end{equation}
where the diffusion pole in the first term is cut by the inelastic
time, $\tau_f$. Substituting $\Gamma ^{\mu \mu }_{\mu \mu }$ and
$\Gamma ^{-\mu \mu }_{-\mu \mu }$ into the general expression
Eq. (\ref{D_1}) and keeping only the terms with coefficients
$\Phi_{\mu \mu}\approx 1$, we arrive at the final result
\begin{equation}
P(\epsilon )   =  -\left( \frac{1}{ 2 \pi E_F \tau _{tr}}
  \right) \left[ \ln \left( \epsilon^2 \tau_{tr}^2 +
    \frac{\tau_{tr}^2}{\tau _{f}^2}  \right) + 
  \ln \left( \epsilon^2 \tau_{tr}^2 + 1 \right) \right].
\label{i}
\end{equation}
Note that each term in the sum (\ref{i}) is comprised of two (equal)
contributions from $\Gamma_{1,1}^{1,1}$  and $\Gamma_{-1,-1}^{-1,-1}$.
In the derivation of Eq.~(\ref{i}), as well as of Eq.~(\ref{diff}),
we used $(v_{\mbox{\tiny F}}\tau_{tr})^{-1}$ as the upper 
cutoff in the integrals over $q$. In principle, there is also a
contribution from the ballistic region, up to 
$q \approx k_{\mbox{\tiny F}}$. This contribution does not
affect the correlation function of the tunneling conductance,
studied in Sec. IV.

\section{Discussion}

The general case of an arbitrary SO-splitting would require a
lengthy calculation, involving $4 \times 4$ diffuson and cooperon
matrices\cite{lyanda98,burkov03}. However, in the most interesting
case of a strong splitting,
when
$\omega_{\mbox{\tiny SO}}\tau_{int}\gg 1$,
a considerable simplification occurs:
the LDOS correlation function, $P(\epsilon )$, at large
enough $\epsilon \gg \tau_{int}^{-1}$ can be expressed in terms
of that in the absence of splitting,
$P(\epsilon ,\omega_{\mbox{\tiny SO}}=0)\equiv P_0(\epsilon )$  The splitting
energy $2\omega_{\mbox{\tiny SO}}$ enters only through the
argument of $P_0$ and the low-energy cutoff $\hbar/\tau_{int}$,
which assumes the role of dephasing time.

To establish the relation between $P$ and $P_0$, we
notice\cite{dmitriev01} that
the states with the same energy, $E$, have two different
values of the wave number,
$\hbar k_{\pm}= \sqrt{2mE}\pm\omega_{\mbox{\tiny SO}}
/v_{\mbox{\tiny F}}$, depending on the branch (see Fig.~1).
Similarly, states at energy $E^{\prime}=E+\epsilon $ have wave numbers
$\hbar k_{\pm}^{\prime}= \sqrt{2mE}+\left(\epsilon
\pm\omega_{\mbox{\tiny SO}}\right)/v_{\mbox{\tiny F}}$. Thus,
we can identify four contributions  to $P(\omega)$, coming from
correlating each of the $k_{\pm}$ states with either of
$k_{\pm}^{\prime}$ states. Since,
for $\omega_{\mbox{\tiny SO}}\tau_{int}\gg 1$,
the degree of correlation between a pair of states depends only
on the difference of the wave numbers, all four contributions
have the same functional form. It is given by the function
$P_0(\epsilon )$ with the corresponding arguments, which can be either
$\hbar v_{\mbox{\tiny F}}(k_{+}^{\prime}- k_{+})=
\hbar v_{\mbox{\tiny F}}(k_{-}^{\prime}- k_{-})=\epsilon $ or
$\hbar v_{\mbox{\tiny F}}(k_{+}^{\prime}- k_{-})=\epsilon +
 2\omega_{\mbox{\tiny SO}}$, or
$\hbar v_{\mbox{\tiny F}}(k_{-}^{\prime}- k_{+})=\epsilon -
2\omega_{\mbox{\tiny SO}}$. It must be emphasized
that, since separate branches cannot be resolved beyond the time
$\tau_{int}$, the correlator $P_0$ for
$\vert \epsilon \vert \approx 2\omega_{\mbox{\tiny SO}}$
must be computed with a low-energy cutoff $\hbar/\tau_{int}$.
Hence the  condition $\epsilon \tau_{int}\gg 1$, adopted in this
qualitative consideration, insures that the SO-induced satellites
in $P(\epsilon )$ do not overlap with the main peak.
Combining the four contributions, we obtain
\begin{equation}
\label{qualitative}
P(\epsilon) = \frac{1}{4}\left[2P_0(\epsilon)+ P_0(\epsilon +
2\omega_{\mbox{\tiny SO}}) +
P_0(\epsilon -2\omega_{\mbox{\tiny SO}})\right].
\end{equation}
Obviously, the above general consideration,  applies to the both
diffuson and cooperon contributions to the LDOS correlator, $P$.
This consideration clarifies why the height of the central  peak
in $P(\epsilon)$ is twice the height of each of the SO satellites.
The factor $2$ in the first term of Eq. (\ref{qualitative})
appears because the  argument $\epsilon$ occurs twice out of the
four possibilities mentioned above.

\section{Manifestation in the Tunneling Spectroscopy}
In the tunneling-spectroscopy experiments
\cite{geim94,savchenko95,falko96,falko97,schmidt97,kuznetsov97,falko00,falko01,falko01'}
the measured quantity
is a resonant-tunneling current, $I_{\mbox{\tiny sd}}$,
through a localized  impurity state as a function of the
source-drain bias,
$V_{\mbox{\tiny sd}}$. Such an impurity plays the role
of a  ``spectrometer'' since its energy position, $\epsilon_0$, changes
with $V_{\mbox{\tiny sd}}$. Within a narrow range, $\delta V$, of
$V_{\mbox{\tiny sd}}$ this change is linear, i.e.,
$\Delta = \epsilon_0(V+\delta V)-\epsilon_0(V)\approx \beta \delta V$,
where $\beta$ is the structure-specific parameter.
Experimentally\cite{geim94,savchenko95,falko96,falko97,schmidt97,kuznetsov97,falko00,falko01,falko01'}, measurements are performed in the
plateau regime when  the temperature is much lower than
$V_{\mbox{\tiny sd}}$, so that the value of  resonant-tunneling
current
\begin{equation}
\label{current}
I(V_{\mbox{\tiny sd}}) = \frac{e}{h} \int_{-\infty }^{\infty}
d\epsilon \frac{\gamma _l (\epsilon ) \gamma _r(\epsilon )}
{[\epsilon -\epsilon_0(V_{\mbox{\tiny sd}})]^2 +\left[
\gamma_l(\epsilon)+ \gamma_r(\epsilon ) \right]^2/4 }
\end{equation}
is temperature independent. In Eq. (\ref{current}), $\gamma_l$
and ${\gamma_r}$ stand for the  tunneling widths, associated
with the escape from the impurity into the emitter and collector,
respectively. These widths are proportional to the LDOS in the
electrodes.
Typically, the widths $\gamma_l$ and $\gamma_r$ differ strongly,
$\gamma_r \gg \gamma_l$. In addition, 
the energy dependence of $\gamma_r$ is weak, so that it can be 
considered as a constant.
As a result, it is a tunneling
coupling to collector that dominates the width of the Lorentzian in Eq.
(\ref{current}). With regard to the weak fluctuating dependence
of current, $I$, on $V_{\mbox{\tiny sd}}$ in the plateau regime,
it is exclusively  due to
the energy dependence of $\gamma_l$, which, in turn, originates
from the fluctuations of the LDOS.
 
Quantitative analysis of the experimental data is performed
 (see e.g. Ref. \onlinecite{falko01}) by plotting the
correlator $C(\delta V)=\langle\delta g(V+\delta V)\delta g(V)\rangle$
of the fluctuations of the  differential conductance
$g=dI_{\mbox{\tiny sd}}/dV$ around its average value,
$\langle g \rangle$,
as a function of $\delta V$ for different values of
$V=V_{\mbox{\tiny sd}}$.
The expression for  $C(\delta V)$ in terms of the LDOS correlator
$P(\epsilon-\epsilon^{\prime})$ is obtained
(see \cite{falko97',falko01''}) by, first, calculating the
the correlator of current fluctuations
$\langle\delta I\left(V_{\mbox{\tiny sd}}\right)
\delta I\left(V^{\prime}_{\mbox{\tiny sd}}\right)\rangle$ and then 
taking derivatives with respect to 
$V_{\mbox{\tiny sd}}$ and
$V^{\prime}_{\mbox{\tiny sd}}$. Using the fact that a convolution of
two lorentzians is also a lorentzian, the final expression
for $C(\delta V)$ can be written in the form
\begin{equation}
\label{tunnel}
C(\delta V)=\langle\delta g(V_{\mbox{\tiny sd}}
+\delta V)\delta g(V_{\mbox{\tiny sd}})\rangle =
- \frac{\langle  \gamma_l \rangle^2 }{4}
   \frac{\partial ^2 }{\partial \Delta ^2 } \left\{
 \gamma_r
  \int  \frac{d\omega \left[P(\omega )
        + P(-\omega ) \right] }
{\left(\omega - \Delta \right)^2 +\gamma_r^2} \right\},
\end{equation}
If, following Ref. \onlinecite{falko97'}, we substitute the
conventional form
$P(\omega)\propto \ln\left(\omega^2\tau_f^2+1\right)$ into
Eq. (\ref{tunnel}), where $\tau_f$ is the ``floating up'' time
of a hole created as a result of the tunneling act, then
Eq.~(\ref{tunnel}) yields\cite{falko97'}
$C(\delta V)= F_2\left(\delta V/V_c\right)$,
where the dimensionless function $F_2(x)$ is defined as
\begin{equation}
F_2(x)=\frac{1-x^2}{\left(1+x^2\right)^2},
\end{equation}
and the characteristic value, $V_c$, is given by
$V_c=\beta^{-1}\left(\gamma_r + \hbar/\tau_{\mbox{\tiny $f$}}\right)$.
In the presence of the SO-coupling the correlator $P(\omega)$
has three peaks, centered at $\omega=0$ and
$\omega=\pm 2\omega_{\mbox{\tiny SO}}$. Without Zeeman splitting
all three peaks have the same shape, which allows to express
the correlator, Eq. (\ref{tunnel}), in terms of the function $F_2$
as follows
\begin{equation}
\label{shifted}
C(\delta V)= \frac{1}{4}F_2\left(\frac{\delta V}{V_c }\right)+
\frac{1}{4}F_2\left(\frac{\delta V}{V_c^{\prime\prime }}\right)
+\frac{1}{4}F_2\left[\frac{\beta\delta V-2\omega_{\mbox{\tiny SO}}}
{\beta V_c^{\prime}}\right] +
\frac{1}{4}F_2\left[\frac{\beta\delta V+
2\omega_{\mbox{\tiny SO}}}{\beta V_c^{\prime}}\right],
\end{equation}
where $V_c^{\prime}=V_c+\beta^{-1}\hbar/\tau_{int}=
\beta^{-1}\left(\gamma_r + \hbar/\tau_{\mbox{\tiny $f$}} +
\hbar/\tau_{int}\right)$,
$V_c^{\prime \prime }=V_c+2\beta^{-1}\hbar/\tau_{int}=
\beta^{-1}\left(\gamma_r + \hbar/\tau_{\mbox{\tiny $f$}} +
2 \hbar/\tau_{int}\right)$,
and $\tau_{int}$ is the intersubband scattering time defined by
Eq.~(\ref{tau5}).
It is seen from Eq. (\ref{shifted}) that the SO peaks in $P(\omega)$
give rise to the satellites in $C(\delta V)$ at
$\delta V=2\beta^{-1}\omega_{\mbox{\tiny SO}}$. Thus, the SO-coupling
has a noticeable effect on the correlator of the tunnel conductances
when  $2\beta^{-1}\omega_{\mbox{\tiny SO}} \geq V_c^{\prime}$.
In Fig. 3 the correlator $C$ is plotted vs. $\delta V/V_c^{\prime}$
for three different values of the dimensionless SO-splitting
$\omega_{\mbox{\tiny SO}}^{\prime}=
\omega_{\mbox{\tiny SO}}/\beta V_c^{\prime}$. It is seen, that
 at $\omega_{\mbox{\tiny SO}}^{\prime}=0.5$ the correlator
$C\left(\delta V/V_c^{\prime}\right)$ develops a characteristic
``shoulder'', while at $\omega_{\mbox{\tiny SO}}^{\prime}=0.7$
it exhibits a well-developed additional maximum. In the experiments
\cite{falko97,falko00,falko01,falko01'} the characteristic width
of the impurity level, $\gamma_r$, was $\lesssim 1K$. This suggests
that the SO satellites should be well resolved in tunneling
spectroscopy of the narrow-gap semiconductors, where
$\omega_{\mbox{\tiny SO}}$ is of the order of several {\em meV}.
Concerning the {\em GaAs}-based structures, the SO splitting
there is much smaller (of the order of $1K$), and depends strongly
on the details of the confinement potential. Therefore, the observation
of the additional peaks in tunneling spectroscopy would require a
higher ``spectrometer resolution''.
The characteristic signature of the SO-related features in 
the correlator $C(\delta V)$ would be a strong sensitivity of 
the correlator to a weak parallel magnetic field. In Fig. 4 we plot
$C\left(\delta V/V_c^{\prime}\right)$ for the same values
$\omega_{\mbox{\tiny SO}}^{\prime}=0.5$ and
$\omega_{\mbox{\tiny SO}}^{\prime}=0.7$ as in Fig. 3 and for different
ratios $\omega_{\mbox{\tiny Z}}/\omega_{\mbox{\tiny SO}}$.
The curves were calculated using Eqs.~(\ref{diff}), (\ref{i})
 for $P(\omega)$. We see that
both the ``shoulder'' at $\omega_{\mbox{\tiny SO}}^{\prime}=0.5$ and
additional maximum at $\omega_{\mbox{\tiny SO}}^{\prime}=0.7$ do
not disappear with increasing $\omega_{\mbox{\tiny Z}}$, but rather
get shifted down. Remarkably, the effect of the parallel field
is noticeable already at very small
$\omega_{\mbox{\tiny Z}}/\omega_{\mbox{\tiny SO}} \sim 0.1$.

\section{Conclusions}
In the present paper we have demonstrated that the intrinsic 
SO-splitting of the electron spectrum in the absence of disorder 
manifests itself in a disorder-induced effect - mesoscopic 
fluctuations of the local density of states. This observation
suggests that the splitting, $2\omega_{\mbox{\tiny SO}}$,
must show up in the resonant tunneling spectroscopy in the 
form of additional peaks in the correlator $C(\delta V)$. 
The fact that the positions of these peaks do not depend 
on the disorder offers a possibility to measure experimentally
the magnitude of the splitting.  Note that, unlike the 
magnetotransport experiments, where the SO-related features 
are quickly washed out with increasing temperature\cite{zum02},
the resonant tunneling current in the plateau regime depends on 
temperature rather weakly\cite{falko00}. As it was demonstrated 
in Sec.\ref{Calculation}, the correlator, $P$, of the LDOS develops
a plateau in a weak parallel magnetic field
$\omega_{\mbox{\tiny Z}} \ll  \omega_{\mbox{\tiny SO}}$.
This singular behavior manifests itself in the anomalous sensitivity 
of the correlator of the tunneling conductances to 
$\omega_{\mbox{\tiny Z}}$, as illustrated in Fig. 4.

\section{Acknowledgements}

This work was  supported by  the NSF Grant No. INT-0231010. B.S.
acknowledges the hospitality of the University of Utah.





\begin{figure}
\centerline{
\epsfxsize=3.4in
\epsfbox{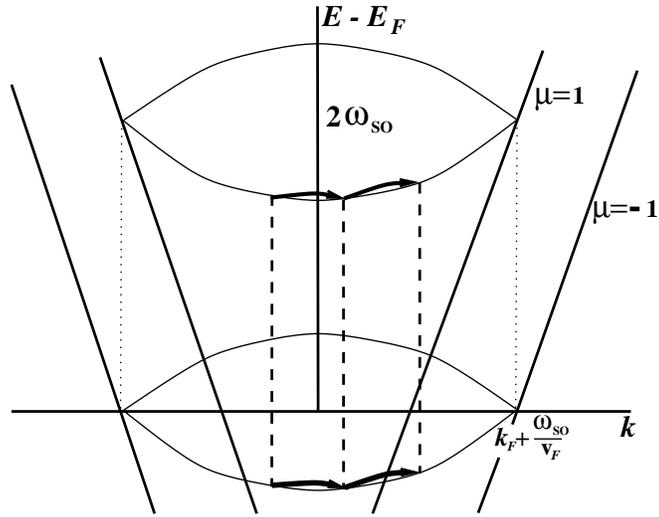}
\protect\vspace*{0.1in}}
\protect\caption[sample]
{\sloppy{ Schematic illustration of the processes responsible
for SO-induced peaks in LDOS correlator at 
$\epsilon = \pm 2 \omega _{\mbox{\tiny SO}}$.
}}
\label{figone}
\end{figure}

\begin{figure}
\centerline{
\epsfxsize=3.4in
\epsfbox{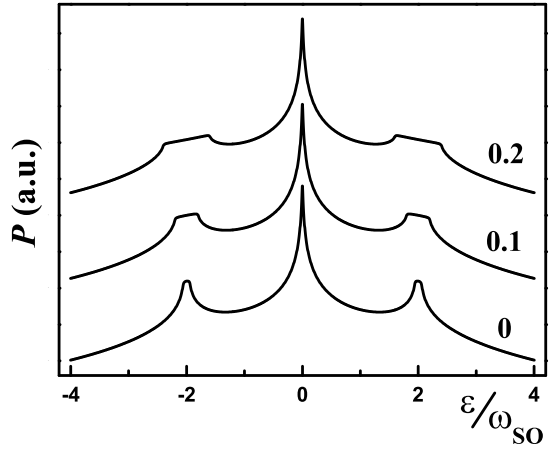}
\vspace*{0.1in}}
\protect\caption[sample]
{\sloppy{ Correlator of the LDOS, $P(\epsilon_1,\epsilon_2)$,
calculated from Eqs.~(\ref{diff}), (\ref{i}) for
$\omega _{\mbox{\tiny SO}}\tau _{tr}=10$  and
$\tau_{f}/\tau _{tr}=10$,
is plotted versus dimensionless energy
$\epsilon / \omega _{\mbox{\tiny SO}}$. The numbers near the lines
show the value of
$\omega _{\mbox{\tiny Z}}/\omega _{\mbox{\tiny SO}}$. For
convenience different curves are shifted along the vertical
axis.
}}
\label{figtwo}
\end{figure}

\begin{figure}
\centerline{
\epsfxsize=3.4in
\epsfbox{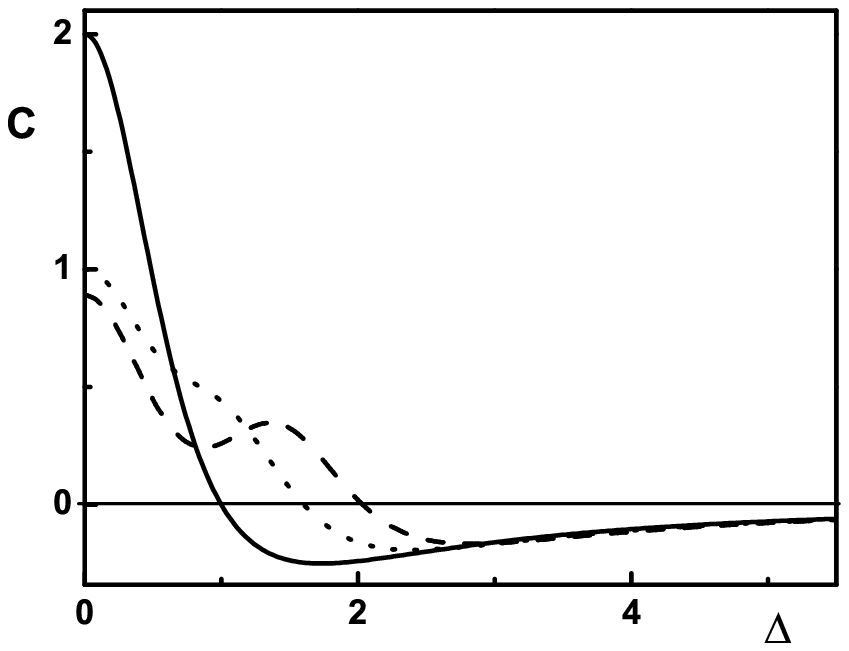}
\vspace*{0.1in}}
\protect\caption[sample]
{\sloppy{The correlator
$C(\delta V)=\langle\delta g(V)\delta g(V+\delta V)\rangle$
of  differential tunneling conductanaces is plotted from
Eq.~(\ref{shifted}) versus
dimensionless voltage difference
$\delta V/V_c^{\prime}$ for different values of the
SO-splitting: $\omega _{\mbox{\tiny SO}}=0$
(solid line); 
$\omega _{\mbox{\tiny SO}}=0.5 V_c^{\prime}$ (dotted line);
$\omega _{\mbox{\tiny SO}}=0.7 V_c^{\prime}$ (dashed line).
For simplicity, 
we have assumed
$\gamma_r \gtrsim \hbar/\tau_{int}$, so that $V_c^{\prime}
\approx V_c \approx V_c^{\prime \prime} $.  
}}
\label{figthree}
\end{figure}

\begin{figure}
\centerline{
\epsfxsize=3.4in
\epsfbox{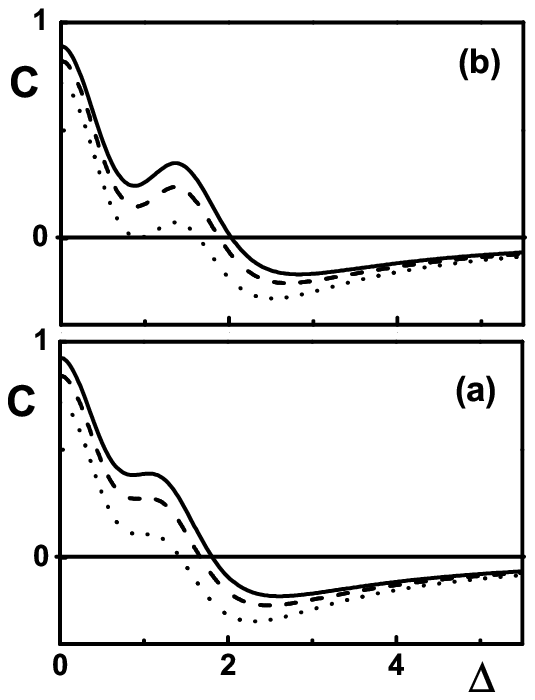}
\vspace*{0.1in}}
\protect\caption[sample]
{\sloppy{Evolution of the SO-related features in
the correlator  $C(\delta V)$
with parallel magnetic field is shown for
same  $\omega _{\mbox{\tiny SO}}=0.5 V_c^{\prime}$ (a)
and $\omega _{\mbox{\tiny SO}}=0.7 V_c^{\prime}$ (b) as in Fig. 3.
Solid lines
$\omega _{\mbox{\tiny Z}}=0$; dahsed lines
$\omega _{\mbox{\tiny Z}}=0.1 \omega _{\mbox{\tiny SO}}$;
dotted lines $\omega _{\mbox{\tiny Z}}=0.2 \omega _{\mbox{\tiny SO}}$.}}
\label{figfour}
\end{figure}


\begin{references}

\vspace{-10mm}

\bibitem{bychkov84} Yu. A. Bychkov and E. I. Rashba,
Pis'ma Zh. Eksp. Teor. Fiz. {\bf 39}, 64 (1984)
[JETP Lett. 39, 78 (1984)].

\bibitem{datta90}S. Datta and B. Das, Appl. Phys. Lett. {\bf 56},
665 (1990).

\bibitem{dorozhkin87} S. I. Dorozhkin and E. B. Ol'shanetskii,
Pis'ma Zh. \'{E}ksp. Teor. Fiz. {\bf 46}, 399 (1987)
[JETP Lett. {\bf 46}, 502 (1987)].

\bibitem{luo90}J. Luo, H. Munekata, F. F. Fang, and P. J. Stiles,
Phys. Rev. B {\bf 38}, 10 142 (1988); Phys. Rev. B {\bf 41}, 7685
(1990).

\bibitem{rudin98}A. M. Rudin, I. L. Aleiner, and L. I. Glazman,
Phys. Rev. B {\bf 58}, 15698 (1998).

\bibitem{raikh93}  M. E. Raikh and T. V. Shahbazyan
Phys. Rev. B {\bf 47}, 1522 (1993).

\bibitem{prigodin91} V. N. Prigodin and M. E. Raikh,
Phys. Rev. B {\bf 43}, 14073 (1991).

\bibitem{lerner92} I. V. Lerner and M. E. Raikh,
Phys. Rev. B {\bf 45}, 14036 (1992).

\bibitem{falko97'} V. I. Fal'ko, Phys. Rev. B {\bf 56}, 1049 (1997).

\bibitem{falko96} T. Schmidt, R. J. Haug, V. I. Fal'ko, K. v. Klitzing,
A. F\"{o}rster, and H. L\"{u}th, Europhys. Lett. {\bf 36}, 61 (1996).

\bibitem{falko97} T. Schmidt, R. J. Haug, V. I. Fal'ko,
K. v. Klitzing, A. F\"{o}rster, and H. L\"{u}th,
Phys. Rev. Lett. {\bf 78}, 1540 (1997).

\bibitem{falko00} J. P. Holder, A. K. Savchenko,
V. I. Fal'ko, B. Jouault, G. Faini, F. Laruelle, and E. Bedel,
Phys. Rev. Lett. {\bf 84}, 1563 (2000).

\bibitem{falko01}J. K\"{o}nemann, P. K\"{o}nig, T. Schmidt,
E. McCann, V. I. Fal'ko, and R. J. Haug, Phys. Rev. B {\bf 64},
155314 (2001).

\bibitem{falko01'}
T. Schmidt, P. K\"{o}nig, E. McCann, V. I. Fal'ko, and R. J. Haug,
Phys. Rev. Lett. {\bf 86}, 276  (2001).

\bibitem{chen99'}G. H. Chen and M. E. Raikh, Phys. Rev. B {\bf 60},
4826 (1999).

\bibitem{prigodin93}V. N. Prigodin, Phys. Rev. B {\bf 47},
10885 (1993).

\bibitem{apalkov02}V. M. Apalkov and M. E. Raikh,
Phys. Rev. Lett. {\bf 89}, 096805 (2002).

\bibitem{lyanda98} Y. Lyanda-Geller,
Phys. Rev. Lett. {\bf 80}, 4273 (1998).

\bibitem{burkov03} A. A. Burkov, and  A. H. MacDonald,
preprint cond-mat/0311328.

\bibitem{dmitriev01}A. P. Dmitriev and V. Yu. Kachorovskii, Phys.
Rev. B {\bf 63}, 113301 (2001).

\bibitem{geim94} A. K. Geim, P. C. Main, N. La Scala, Jr., L. Eaves,
T. J. Foster, P. H. Beton, J. W. Sakai, F. W. Sheard,  M. Henini,
G. Hill and M. A. Pate, Phys. Rev. Lett. {\bf 72}, 2061 (1994).

\bibitem{savchenko95}A. K. Savchenko, V. V. Kuznetsov,
A. Woolfe, D. R. Mace, M. Pepper, D. A. Ritchie, and G. A. C. Jones
       Phys. Rev. B {\bf 52}, R17021 (1995).

\bibitem{schmidt97} T. Schmidt, R. J. Haug, K. v. Klitzing,
A. F\"{o}rster, and H. L\"{u}th, Phys. Rev. B {\bf 55}, 2230 (1997).

\bibitem{kuznetsov97}V. V. Kuznetsov, A. K. Savchenko, D. R.
Mace, E. H. Linfield, and D. A. Ritchie,
       Phys. Rev. B {\bf 56}, R15533 (1997).

\bibitem{falko01''} E. McCann and V. I Fal'ko,
J. Phys.: Condens. Matter {\bf 13} 6633 (2001).

\bibitem{zum02} D. M. Zumb\"{u}hl, J. B. Miller, C. M. Marcus, K. Campman, and A. C. Gossard
 Phys. Rev. Lett. {\bf 89}, 276803 (2002).




\end{references}
\end{document}